\documentclass[showpacs,aps,twocolumn]{revtex4}
\newcommand{\bra}[1]{\langle #1|}
\newcommand{\ket}[1]{|#1\rangle}

\def\Pro{{\sf P}}

\newcommand{\tr}{{\rm tr}}

\newcommand{\e}{{\rm e}}

\newcommand{\omegav}{\boldsymbol{\omega}}

\newcommand{\p}{{\rm p}}
\newcommand{\x}{{\rm x}}

\newcommand{\omt}{\frac{\omega}{T}}
\newcommand{\chiomt}{\chi\!\left( \omt \right)}

\usepackage{epsfig}
\usepackage{graphicx}
\usepackage{amsmath,amssymb,amsfonts}
\usepackage{array}
\usepackage{url}
\usepackage{multirow}
\usepackage{natbib}
\usepackage{float}
\usepackage{lineno}
\usepackage{xspace}
\usepackage{ulem}

\newcommand{\bef}{\begin{figure}}
\newcommand{\eef}{\end{figure}}
\newcommand{\bc}{\begin{center}}
\newcommand{\ec}{\end{center}}

\newcommand{\be}{\begin{equation}}
\newcommand{\ee}{\end{equation}}
\newcommand{\bea}{\begin{eqnarray}}
\newcommand{\eea}{\end{eqnarray}}

\def\ba{\begin{eqnarray}}
\def\ea{\end{eqnarray}}
\usepackage[usenames,dvipsnames]{color}
\definecolor{darkblue}{RGB}{0,0,196}
\usepackage[colorlinks=true,linkcolor=darkblue,citecolor=darkblue,urlcolor=darkblue]{hyperref}

\begin{document}
\title{Thermodynamics of a rotating hadron resonance gas with van der Waals interaction}
 
\author{Kshitish Kumar Pradhan}
\author{Bhagyarathi Sahoo}
\author{Dushmanta Sahu}
\author{Raghunath Sahoo\footnote{Corresponding author: Raghunath.Sahoo@cern.ch}}
\affiliation{Department of Physics, Indian Institute of Technology Indore, Simrol, Indore 453552, India}

\begin{abstract}

Studying the thermodynamics of the systems produced in ultra-relativistic heavy-ion collisions is crucial in understanding the QCD phase diagram. Recently, a new avenue has opened regarding the implications of large initial angular momentum and subsequent vorticity in the medium evolution in high-energy collisions. This adds a new type of chemical potential into the partonic and hadronic systems, called the rotational chemical potential. We study the thermodynamics of an interacting hadronic matter under rotation, formed in an ultra-relativistic collision. We introduce attractive and repulsive interactions through the van der Waals equation of state. Thermodynamic properties like the pressure ($P$), energy density ($\varepsilon$), entropy density ($s$), trace anomaly ($(\varepsilon - 3P)/T^{4}$), specific heat ($c_{\rm v}$) and squared speed of sound ($c_{\rm s}^{2}$) are studied as functions of temperature ($T$) for zero and finite rotation chemical potential. The conserved charge fluctuations, which can be quantified by their respective susceptibilities, are also studied. The rotational (spin) density corresponding to the rotational chemical potential is explored. In addition, we explore the possible liquid-gas phase transition in the hadron gas with van der Waals interaction in the $T$ -- $\omega$ phase space.

\end{abstract}
\date{\today}
\maketitle

\section{Introduction}
\label{intro}

There have been intense investigations to understand the behavior of the strongly interacting matter produced in the ultra-relativistic heavy-ion collisions at the Relativistic Heavy Ion Collider (RHIC) at BNL and the Large Hadron Collider (LHC) at CERN. Such matter can be described by Quantum Chromodynamics (QCD). According to QCD, a smooth crossover phase transition is expected at high temperatures ($T$) and vanishing baryochemical potentials ($\mu_{\rm B}$) region in the QCD phase diagram, which is described by the RHIC and LHC experiments. As one goes towards low temperatures, and high baryochemical potential region, the phase transition becomes a first-order one. These phase transition lines meet at a hypothesized critical endpoint (CEP), which has been one of the most exciting topics of discussion in the high-energy physics community. To gather valuable and reliable information about the QCD matter, lattice QCD (lQCD) has been the most successful theory based on first principles. However, at non-vanishing chemical potential, the lQCD breaks down because of the fermion sign problem \cite{Borsanyi:2013bia, HotQCD:2014kol}. Nevertheless, there have been significant attempts to bypass this problem indirectly \cite{Allton:2005gk,deForcrand:2002hgr, Fodor:2001pe}, but the issue still largely persists. The more simplistic Hadron Resonance Gas (HRG) model effectively explains the QCD matter behavior and matches the lQCD data up to temperature ($T \simeq 150$ MeV) \cite{Bellwied:2013cta, HotQCD:2012fhj, Bellwied:2017ttj}. The HRG model works at both zero baryochemical potential and baryon-rich environments and successfully explains the hadron yields from heavy-ion collisions using only two parameters, the temperature, and the baryochemical potential. However, after $T \simeq 150$ MeV, the hadrons start to melt, and the results from the HRG model start deviating from the lQCD estimations. The HRG model also breaks down while estimating the higher-order fluctuations and correlations of conserved charges \cite{Bazavov:2013dta}. It has been observed that repulsive interactions between the hadrons can substantially affect the behavior of thermodynamic and transport properties, particularly the higher order fluctuations \cite{Pal:2020ucy, Bhattacharyya:2013oya}. The repulsive interactions can be incorporated into the hadron gas by a van der Waals type repulsion where the hardcore radius of the hadrons serves as the repulsion in the Excluded Volume Hadron Resonance Gas (EVHRG) model or through a repulsive mean field potential which is introduced in the hadron gas in the Relativistic Mean-Field Hadron Resonance Gas (RMFHRG) model. Recently, an interacting hadron resonance gas model was introduced where both long-distance attraction and short-distance repulsion between the hadrons was taken into account through the van der Waals equation of state \cite{Vovchenko:2016rkn}. The van der Waals-type interaction in the hadronic medium delays the melting of hadrons, and thus, the VDWHRG model can explain the lQCD data even up to $T \sim$ 180 MeV. This model, which has a liquid-gas phase transition, is very effective in estimating a variety of thermodynamic and transport properties \cite{Pradhan:2022gbm, Sarkar:2018mbk, Samanta:2017yhh, Sahoo:2023vkw}. 

In order to understand the medium formed in ultra-relativistic collisions, it is essential to study its thermodynamic properties. The fundamental thermodynamic quantities, such as pressure ($P$), energy density ($\varepsilon$), and entropy density ($s$), can give us necessary information about the system. The scaled pressure, energy density, and entropy density provide information about the degrees of freedom in the medium. Similarly, the speed of sound tells about the interaction in the medium. A massless ideal gas gives $c_{\rm s}^{2}$ to be 1/3, whereas the value for a hadron gas is 1/5. Studying $c_{\rm s}^{2}$ can help us understand whether the medium is partonic or hadronic or approaches a massless ideal gas limit. On the other hand, the specific heat of a system is estimated via the temperature fluctuations in the system. It gives the measure of the amount of heat energy needed to raise the system's temperature by one unit. It is also expected to diverge near the critical point and thus is an excellent observable to study the phase transition. Similarly, trace anomaly also plays a vital role in the QCD dynamics and phase transition. It measures the deviation from the masslessness of the constituents in the medium. In recent studies \cite{Sahu:2020nbu, Sahu:2020swd, Sahu:2020mzo, Sahu:2021cdl}, various thermodynamic observables were studied as functions of final state charged particle density in pseudorapidity ($\langle dN_{\rm ch}/d\eta \rangle$), which shed light on a possible change in dynamics of the system after a threshold in $\langle dN_{\rm ch}/d\eta \rangle$. This suggests that after $\langle dN_{\rm ch}/d\eta \rangle \simeq$ 10 -- 20, small systems like pp and p-Pb mimic the behavior of heavy-ion collisions. 

In addition, studying the correlations and fluctuations of conserved charges is a reliable method for comprehending the physics of the phase transition of strongly interacting matter. As the fluctuation-dissipation theorem relates susceptibilities to fluctuations in a system near thermal equilibrium, the related susceptibilities indicate inherent statistical fluctuations. Changes in conserved charges at finite temperatures and chemical potential are sensitive signs that hadronic matter is changing into quark-gluon plasma (QGP). Moreover, divergent fluctuations can also indicate the presence of the CEP; however, the shift from the hadronic to QGP phase is continuous for the vanishing net baryon chemical potential.

Recently, it was found that there is a finite hyperon polarization in relativistic heavy-ion collisions at the STAR experiment, which led to the conclusion that finite vorticity is present in the medium \cite{STAR:2017ckg}. This opens up a whole new window of exciting consequences. The vorticity or rotation gets coupled with the temperature in the medium, thus changing the entire dynamics of the system evolution \cite{Singh:2018bih, Sahoo:2023xnu}. The fundamental Euler's thermodynamic equation gets modified in the presence of a finite rotation, adding a new rotation chemical potential into the system ~\cite{Becattini:2010, Florkowski:2017ruc}. Apart from the vorticity coming from the initial global orbital angular momentum of the colliding heavy ions, there are several other sources from which vorticity can be generated in a system. The smoke-loop type vortex created in the vicinity of fast-moving jets in an expanding fireball also contributes to the vorticity of the system, although they are not responsible for hyperon polarization ~\cite{Betz:2007kg}. Inhomogeneous transverse expansion of fireball may produce transverse vorticity circling the longitudinal axis in the system ~\cite{Xia:2018tes, Jiang:2016woz, Wei:2018zfb, Becattini:2017gcx, Pang:2016igs, Voloshin:2017kqp}. Additionally, vorticity can be generated from the Einstein-de Haas effect ~\cite{Einstein:1915}, where a magnetized medium creates a finite rotation. This indicates that the huge magnetic field produced in heavy ion collisions due to fast-moving spectators may magnetize the medium and it may generate a large amount of vorticity in the system, whereas the reverse effect is the famous Barnett effect~\cite{Barnett}. Thus, the analogy between rotation and the magnetic field is a well-known phenomenon studied in many physical systems~\cite{McInnes:2016dwk,Yang:2023zqe}. Similarly, viscosity in the medium is also responsible for generating finite vorticity and vice versa ~\cite{Fu:2021pok,Sahoo:2023xnu}. Thus, it is necessary to include the effect of rotation while studying the medium formed in an ultra-relativistic collision. Recently, many studies have been conducted with the introduction of rotation to understand the QCD phase structure. Vorticity formation in the ultra-relativistic heavy-ion collision has been studied from hydrodynamic models such as ECHO-QGP, PICR, vHLLE, MUSIC, 3-FD, CLVisc in (3+1) dimensional model~\cite{Becattini:2015ska,Csernai:2013bqa, Csernai:2014ywa, Ivanov:2019ern, Karpenko:2016jyx}. Event generators, such as AMPT, UrQMD, and HIJING, have also been used to estimate kinematic and thermal vorticity~\cite{Jiang:2016woz, Deng:2016gyh, Li:2017slc, Wei:2018zfb, Deng:2020ygd, Vitiuk:2019rfv}. Moreover, the non-zero local vorticity can help us probe the chiral vortical effect (CVE), which is a non-trivial consequence of topological quantum chromodynamics~\cite{Rogachevsky:2010ys, Kharzeev:2007tn}. This effect is the vortical analog of the chiral magnetic effect (CME)~\cite{Kharzeev:2007jp, Fukushima:2008xe} and chiral separation effect (CSE) \cite{Son:2004tq, Metlitski:2005pr}. It represents the vector, and axial currents generation along the vorticity ~\cite{Banerjee:2008th, Erdmenger:2008rm, Son:2009tf, Jiang:2015cva}. CVE is extremely important because it induces baryon charge separation along the vorticity direction, which can be experimentally probed by two-particle correlations~\cite{Csernai:2013vda}.

There are several studies on the effect of magnetic fields on the QCD phase diagram. In \cite{Mizher:2010zb}, the authors have coupled the linear sigma model to quarks to study the chiral transition as well as to Polyakov loops to consider the confinement. Taking a constant uniform magnetic field, they investigated how the magnetic field affects the chiral and deconfinement transitions. It was shown that the chiral condensate is enhanced by the magnetic field, and the transition temperature rises as a result. This phenomenon is commonly known as magnetic catalysis. Numerous studies that involve the Nambu-Jona-Lasinio model and its extended versions, such as the PNJL \cite{Kashiwa:2011js}, and EPNJL \cite{Gatto:2010pt} models came to similar conclusions regarding the rise in $T_{c}$ and the strength of the transitions. However, in contrast to this, the lattice QCD results \cite{Bali:2011qj, Bali:2012zg}, showed that the magnetic field actually suppresses rather than increases the critical temperature for the chiral phase transition of QCD. Since rotation in the medium adds another kind of chemical potential, it can affect the phase transition, and hence, it will be intriguing to observe how rotation affects the QCD phase diagram. In ref. \cite{Fujimoto:2021xix}, the authors explore the deconfinement from a rapidly rotating hot and dense hadronic matter, similarly, in ref. \cite{Jiang:2016wvv,Wang:2018sur}, the authors investigate the chiral phase transition in a system of fermions under rotation using the Nambu-Jona Laisino (NJL) model. Here the authors have shown the importance of angular velocity in determining the phase transition from hadronic to quark degrees of freedom. Their results are presented by a phase diagram in the temperature-angular momentum $T-\omega$ plane along with the phase diagram in the temperature-chemical potential $T-\mu_{B}$ plane.
Recent studies show that the rotational motion of QGP may result in a negative moment of inertia and hence becomes unstable \cite{Braguta:2023yjn}. Moreover, a rotating plasma can produce inhomogeneity in the medium \cite{Chernodub:2020qah,Chernodub:2022veq}. However, for a system of homogenous hadronic matter, this instability can be ignored. Similarly, the quark deconfinement has been studied in rotating neutron stars in \cite{Mellinger:2017dve}. The authors in ref. \cite{Sun:2021hxo} have studied the chiral phase transition along with the spin polarization in a three-flavor NJL model. Apart from this, the rotation effect has been explored in the mesonic condensation of the isospin matter \cite{Zhang:2018ome}. The authors in \cite{Liu:2017spl} have also studied a combined effect of rotation and magnetic field on pion condensation. They have demonstrated increased condensation upon increasing the rotation (angular velocity). These kinds of studies suggest that introducing angular velocity in the medium adds another kind of chemical potential known as rotational chemical potential. 
Similar to baryon chemical potential, this rotational chemical potential can also lead to a phase transition. 

Given the above information, it would be interesting to understand what effect rotation plays in quantifying a hadronic system's thermodynamic properties. In this work, we take rotation into account and estimate various thermodynamic properties and charge fluctuations in an interacting hadron gas for the first time. We also look for a possible criticality in the rotating medium for a liquid-gas phase transition. The structure of this work is as follows. The section \ref{formulation} gives a detailed calculation of the thermodynamic observables and the susceptibilities within the scope of a VDWHRG model that includes rotation. We briefly examine the results in section \ref{results} and provide a summary in section \ref{sum}.

\section{Formulation}
\label{formulation}
In this work, we have assumed a system of relativistic gas of massive fermions and bosons having half-integral and integral spin ($S$), rotating with a constant angular velocity vector $\omega$. The density operator for rotational grand canonical ensemble having large  volume $V$ and angular momentum  ${\bf J}$ is given as ~\cite{Becattini:2007nd, Becattini:2009wh, landau, beca1, beca2, Becattini:2013fla}

\begin{equation}
\label{denmat2}
 \widehat\rho_{\omega} = \frac{1}{Z_\omega} \exp [ (-\widehat H + \mu \widehat Q +  {\bf \omega} \cdot \widehat {\bf J})/T] \Pro_V,
\end{equation}
where $\widehat H $  is the Hamiltonian operator, $\widehat Q $ is a generic conserved charges, $\mu $ is the relevant chemical potential, and $Z_\omega$ is the partition function of the rotating system given as
\begin{equation}\label{zomega}
  Z_\omega = \tr \exp [ (-\widehat H + \mu \widehat Q + 
  \omega \cdot \widehat {\bf J})/T] \Pro_V.
\end{equation}

The $\Pro_V $ is the projector onto  localized states $\ket{h_V}$, $  \Pro_V = \sum_{h_V} \ket{h_V}\bra{h_V}$.
The partition function in Eq.~(\ref{zomega}) can be written as a product of single-particle partition functions. The calculation of Eq.~(\ref{zomega}) then reduces to compute matrix elements of the single-particle Hamiltonian, charge, and angular momentum compatible operators $\widehat h,\widehat q, \widehat {\bf j}$, respectively like this ~\cite{Becattini:2007nd, Becattini:2009wh}:

\begin{equation}\label{matel1}\begin{split}
  &\bra{p_{\pm},\tau_{\pm}} \exp[-(\widehat h + \mu \widehat q + \omega
  \cdot \widehat {\bf j})/T] \Pro_V \ket{p_{\pm}, \sigma_{\pm}}\\
  &= (\e^{(\epsilon - \mu q )/T} \pm 1)^{-1} \bra{p_{\pm},\tau_{\pm}} \exp[\omega 
  \cdot \widehat {\bf j}/T] \Pro_V \ket{p_{\pm}, \sigma_{\pm}},
\end{split}  
\end{equation}
where $p_{\pm}$ is the single particle four-momentum. The $\sigma_{\pm}$ and $\tau_{\pm}$ are level of  polarization states. The $\pm$  corresponds to fermions and bosons, respectively. An analytical extension from imaginary values of $\omega$ can be used to derive the matrix elements on the right-hand side of Eq.~(\ref{matel1}). After replacing $\omega/T$ with $-i \phi$, we have a rotation ${\sf R}_{\hat \omega}(\phi)$
around the axis $\omega$ of an angle $\phi$. A more detailed explanation can be found in the Ref. ~\cite{Becattini:2007nd, Becattini:2009wh}. 
\begin{widetext}    
The matrix element can be rewritten in terms of the rotational matrix ${\sf R}_{\hat \omega}(\phi)$ as 
\begin{equation}\label{matel1.1}
    \bra{p_{\pm},\tau_{\pm}} \exp[\omega 
  \cdot \widehat {\bf j}/T] \Pro_V \ket{p_{\pm}, \sigma_{\pm}}  = \bra{p_{\pm},\tau_{\pm}} {\sf R}_{\hat \omega}(\phi) \Pro_V \ket{p_{\pm}, \sigma_{\pm}}. 
  \end{equation}
  Expanding the matrix element of the right-hand side,  
\begin{equation}\label{matel2}
  \bra{p_{\pm},\tau_{\pm}} {\sf R}_{\hat \omega}(\phi) \Pro_V \ket{p_{\pm}, \sigma_{\pm}} 
   =   \sum_{\sigma'_{\pm}} \int d^3 \p'  \; \bra{p_{\pm},\tau_{\pm}} \widehat{\sf R}_{\hat \omega} 
   (\phi) \ket{p'_{\pm},\sigma'_{\pm}} \bra{p'_{\pm},\sigma'_{\pm}} \Pro_V \ket{p_{\pm}, \sigma_{\pm}}. 
\end{equation}
The matrix element representation of the rotation involves a Dirac delta and a Wigner matrix. Thus,
\begin{equation}\label{rota}
\bra{p_{\pm},\tau_{\pm}} \widehat{\sf R}_{\hat \omega}(\phi)\ket{p'_{\pm},\sigma'_{\pm}}=
 \delta^3\left({\bf p_{\pm}} - {\sf R}_{\hat \omega}(\phi)({\bf p_{\pm}}')\right) D^S([{\sf R}_{\hat \omega}(\phi)(p'_{\pm})]^{-1} {\sf R}_{\hat \omega}(\phi)[p'_{\pm}])_{\tau_{\pm} \sigma'_{\pm}}.
\end{equation}
It's difficult to determine the $\Pro_V$ matrix element over momentum eigenstates. A theoretical quantum field framework can be used to perform the calculation~\cite{beca2}%
\begin{equation}\label{prov} 
  \bra{p'_{\pm},\sigma'_{\pm}} \Pro_V \ket{p_{\pm},\sigma_{\pm}} = \frac{1}{2}
  \sqrt{\varepsilon \over \varepsilon'} \, \int_V d^3 \x \; \e^{i {\bf x} \cdot ({\bf p_{\pm}}-{\bf p_{\pm}}')} \left( D^S([p'_{\pm}]^{-1}[p_{\pm}])+D^S([p'_{\pm}]^{\dagger}[p_{\pm}]^{\dagger-1})\right)_{\sigma'_{\pm}\sigma_{\pm}} \bra{0} \Pro_V \ket{0},
\end{equation} 
 where $\bra{0} \Pro_V \ket{0} $ is the vaccum expectation of the projector $\Pro_{V}$ and tends to 1 for large volume. Substituting the Eqs. (\ref{rota}) and (\ref{prov}) in Eq. (\ref{matel2}), we have 
 
 \begin{equation}\label{matel3}
  \bra{p_{\pm},\tau_{\pm}} \widehat{\sf R}_{\hat \omega}(\phi) \Pro_V \ket{p_{\pm}, \sigma_{\pm}} =
  \int_V d^3 \x \; \e^{i {\bf x} \cdot 
  ({\bf p_{\pm}}-{\sf R}_{\hat \omega}(\phi)^{-1}({\bf p_{\pm}}))} \;\; 
  \frac{1}{2} \left( D^S([p_{\pm}]^{-1} {\sf R}_{\hat \omega}(\phi)[p_{\pm}])+
  D^S([p_{\pm}]^{\dagger} {\sf R}_{\hat \omega}(\phi) [p_{\pm}]^{\dagger-1}) 
  \right)_{\tau_{\pm} \sigma_{\pm}}
\end{equation}  
Taking advantage of the unitarity of the Wigner rotation, i.e.,
\begin{equation}\label{unitary}
 D^S([{\sf R}_{\hat \omega}(\phi)(p'_{\pm})]^{-1} {\sf R}_{\hat \omega}(\phi)
 [p'_{\pm}]) =  D^S([{\sf R}_{\hat \omega}(\phi)(p'_{\pm})]^\dagger 
 {\sf R}_{\hat \omega}(\phi) [p'_{\pm}]^{\dagger -1})
\end{equation} 
and the unitarity of ${\sf R}$ itself as an SL(2,C) matrix.
The analytical prolongation of equation (\ref{matel3}) to imaginary angles
yields the final expression for the matrix element in Eq. (\ref{matel1}) as:
\begin{equation}\label{matel4}\begin{split}
  \bra{p_{\pm},\tau_{\pm}} \exp[\omega \cdot \widehat {\bf j}/T] \Pro_V \ket{p_{\pm}, \sigma_{\pm}} =&
  \int_V d^3 \x \; \e^{i {\bf x} \cdot 
  ({\bf p_{\pm}}-{\sf R}_{\hat \omega}(i \omega/T)^{-1}({\bf p_{\pm}}))}\\
  &\times \frac{1}{2} \left( D^S([p_{\pm}]^{-1} {\sf R}_{\hat \omega}(i \omega/T)[p_{\pm}])
  +D^S([p_{\pm}]^{\dagger} {\sf R}_{\hat \omega}(i \omega/T) [p_{\pm}]^{\dagger-1}) 
  \right)_{\tau_{\pm} \sigma_{\pm}}
  \end{split}
\end{equation}
The equilibrium single-particle phase space distribution can be calculated with the help of the matrix element in Eq. (\ref{matel4}). The spacial integral form in Eq. (\ref{matel1}) allows us to write the phase-space distribution as:
\begin{equation}\label{phsp}
  \begin{split}
  f({\bf x},{\bf p})_{\tau_{\pm} \sigma_{\pm}} = &(e^{(\varepsilon-\mu q)/T} \pm 1)^{-1}
  \e^{i {\bf x} \cdot  ({\bf p_{\pm}}-{\sf R}_{\hat \omega}(i \omega/T)^{-1}({\bf p_{\pm}}))}\\   
  &\times \frac{1}{2} \left( D^S([p_{\pm}]^{-1} {\sf R}_{\hat \omega}(i \omega/T)[p_{\pm}]) + D^S([p_{\pm}]^{\dagger} {\sf R}_{\hat \omega}(i \omega/T) [p_{\pm}]^{\dagger-1}) \right)_{\tau_{\pm} \sigma_{\pm}}
  \end{split}
\end{equation}

A non-relativistic thermodynamic equilibrium rotating system is possible when the rotation is rigid \cite{Becattini:2007nd, landau}. That is, for a system of size $x$, we can define a constant angular velocity vector, $\omega$, so as to have the rigid velocity, $\bold{v} = \omegav\times \bold{x}$, which in the relativistic system adds another constraint of $|\omegav\times \bold{x}| \ll 1$ in natural units. Therefore the ratio between $\omega$ and $T$ is very small for a proper macroscopic system (and in fact, for the majority of practical reasons), i.e.: $\frac{\hbar \omega}{k_{B}T} \ll 1$. 
As a result, the lowest order term in $\omega/T$ is a good approximation for the difference between the momenta in the exponent of Eq. (\ref{phsp}). 

\begin{equation}\label{vectdiff}
 {\bf p_{\pm}} - {\sf R}_{\hat \omega}(i \omega/T)^{-1}({\bf p_{\pm}}) = {\bf p_{\pm}} - 
 \left[ \cosh \omt \, {\bf p_{\pm}} - i \sinh \omt \, \hat \omega \times {\bf p_{\pm}} + 
 (1-\cosh \omt) \, {\bf p_{\pm}}\cdot \hat \omega \hat \omega \right] \simeq
 i \omt \hat \omega \times {\bf p_{\pm}}
\end{equation}
This results in the phase-space distribution function in Eq. (\ref{phsp}) becoming;
\begin{eqnarray}\label{phsp2}
  f({\bf x},{\bf p_{\pm}})_{\tau_{\pm}\sigma_{\pm}} &=&  (e^{(\varepsilon - \mu q)/T} \pm 1)^{-1} 
 e^{ - {\bf x} \cdot (\omega \times {\bf p_{\pm}})/T }
  \frac{1}{2} \left( D^S([p_{\pm}]^{-1} {\sf R}_{\hat \omega}(i \omega/T)[p_{\pm}])+
  D^S([p_{\pm}]^{\dagger} {\sf R}_{\hat \omega}(i \omega/T) [p_{\pm}]^{\dagger-1}) 
  \right)_{\tau_{\pm} \sigma_{\pm}} \nonumber \\
  &=& (e^{(\varepsilon - \mu q)/T} \pm 1)^{-1} e^{ {\bf p_{\pm}} \cdot (\omega \times {\bf x})/T }\frac{1}{2} \left( D^S([p_{\pm}]^{-1} {\sf R}_{\hat \omega}
  (i \omega/T)[p_{\pm}])+ D^S([p_{\pm}]^{\dagger} {\sf R}_{\hat \omega}(i \omega/T) 
  [p_{\pm}]^{\dagger-1}) \right)_{\tau_{\pm} \sigma_{\pm}}
  \nonumber \\
  &=&  (e^{(\varepsilon - \mu q)/T} \pm 1)^{-1}  e^{ ({\bf p_{\pm}} \cdot \bf v)/T }\frac{1}{2} \left( D^S([p_{\pm}]^{-1} {\sf R}_{\hat \omega}
  (i \omega/T)[p_{\pm}])+ D^S([p_{\pm}]^{\dagger} {\sf R}_{\hat \omega}(i \omega/T) 
  [p_{\pm}]^{\dagger-1}) \right)_{\tau_{\pm} \sigma_{\pm}}
\end{eqnarray}
where we have used the definition of ${\bf v} = \boldsymbol{ \omega} \times \bf{x}$
\end{widetext}

The single-particle phase-distribution in Eq. (\ref{phsp2}) for the ideal rotating relativistic fermions and bosons particles is the unnormalized one, and we need to take the trace of the matrix in Eq. (\ref{phsp2}) to obtain the so-called phase-space density in $({\bf x},{\bf p})$:
\begin{equation}\label{phsp1}\begin{split}
  f({\bf x},{\bf p}) &= \sum_{\sigma_{\pm}} f({\bf x},{\bf p})_{\sigma_{\pm} \sigma_{\pm}}\\
  &=(e^{(\varepsilon - \mu q)/T} \pm 1)^{-1}  e^{ ({\bf p} \cdot \bf v)/T } \chiomt,
\end{split}
\end{equation}  
being:
\begin{equation}\label{chiomegat}
\chiomt \equiv \tr D^S({\sf R}_{\hat \omega}(i \omega/T)) =
 \frac{\sinh(S+\frac{1}{2})\omt}{\sinh(\frac{\omega}{2T})}  ,
\end{equation}
where the q is the conserved charge.

\subsection{van der Waals HRG model}
\label{ss1}

In contrast to the QGP phase, where the degrees of freedom are basically quarks and gluons, the hadronic phase is described by the confined state of the quarks and gluons. The ideal HRG model deals with a system of non-interacting point particles with hadronic degrees of freedom. The thermodynamic pressure for $i^{th}$ particle is given by ~~\cite{Vovchenko:2015vxa, Vovchenko:2015pya},
\begin{equation}
\label{pressure}
P^{id}_i = \frac{g_i}{2\pi^2} \int_{0}^{\infty} p^2 dp\ \frac{p^2}{3E_i}\frac{1}{\exp[(E_i - \mu_i)/T] \pm 1}.
\end{equation}
Here, the degeneracy of ${i^{th}}$ hadronic species is given by $g_i$, whereas $E_i = \sqrt{p^2 + m_i^2}$ gives the free particle energy of the $i^{th}$ hadron, $m_{i}$ being the mass of the ${i^{th}}$ hadron. The $\pm$ sign corresponds to baryons and mesons, respectively.  The chemical potential is denoted by $\mu_{i}$, and is given by,
\begin{equation}
\label{mub}
\mu_i = B_i\mu_B + S_i\mu_S +Q_i\mu_Q,
\end{equation}
where $\mu_{B}$, $\mu_{S}$, and $\mu_{Q}$, respectively, represent the baryon chemical potential, strangeness chemical potential, and charge chemical potential. The baryon number, strangeness, and electric charge of the ${i^{th}}$ hadron are denoted by $B_i$, $S_i$, and $Q_i$, respectively.

In a rotating medium of hadron gas, the pressure for a single hadronic species is equivalent to the one defined in Eq. (\ref{pressure}) multiplied by a factor $e^{ ({\bf p} \cdot \bf v)/T }\chi (\frac{\omega}{T})$, given as
\begin{equation}
\begin{split}
\label{rot_part}
P^{id}_i = \frac{g_i}{2\pi^2} \int_{0}^{\infty} p^2 dp \frac{p^2}{3E_i}\frac{e^{ ({\bf p} \cdot \bf v)/T}}{\exp[(E_i - \mu_i)/T] \pm 1}\chiomt,
\end{split}
\end{equation}
where $\chi (\frac{\omega}{T})$ is given by the Eq. (\ref{chiomegat}). Respecting the causality condition $\omegav\times \bold{x} \ll 1$, we can safely neglect the $x$ dependency and fix the system size, $x=R$, to be 5 fm throughout the work. Also, for the values of $\omega$ we have used in this study, the Lorentz factor is close to unity and will have a negligible impact on the results. Hence, it is neglected in the current formalism. The total pressure of the hadron gas is then obtained by summing over the pressure of all hadron species as; 

\begin{equation}
    \label{total}
    P = \sum_{i} P_{i}^{id}.
\end{equation}
Different thermodynamic quantities for a single hadronic species, such as energy density $\varepsilon_i$, number density $n_i$, and entropy density $s_i$, can be written as
\begin{equation}
\label{energy}
\varepsilon^{id}_i(T,\mu_i) = \frac{g_i}{2\pi^2} \int_{0}^{\infty} \frac{E_i\ p^2 dp\ e^{ ({\bf p} \cdot \bf v)/T }}{\exp[(E_i-\mu_i)/T]\pm1}\chi (\frac{\omega}{T})
\end{equation}
\begin{equation}
\label{numden}
n^{id}_i(T,\mu_i) = \frac{g_i}{2\pi^2} \int_{0}^{\infty} \frac{p^2 dp\ e^{ ({\bf p} \cdot \bf v)/T }}{\exp[(E_i-\mu_i)/T]\pm1}\chi (\frac{\omega}{T})
\end{equation}
\begin{align}
 s^{id}_i(T,\mu_i)=&\pm\frac{g_i}{2\pi^2} \int_{0}^{\infty} p^2 dp\ \frac{p^2}{3E_i}\frac{e^{ ({\bf p} \cdot \bf v)/T}}{\exp[(E_i-\mu_i)/T]\pm1} \nonumber\\ 
 &\times \Bigg[\frac{1}{T^2}\Big(\frac{(E_i - \mu_i)}{1\pm\exp[-(E_i-\mu_i)/T]} - {\bf p} \cdot \bf v\Big)\chi (\frac{\omega}{T}) \nonumber\\ 
&+ \frac{\partial}{\partial T}\chiomt \Bigg]
 \label{entropy}
 \end{align}
 Along with this, we also compute another density related to angular velocity (the so-called rotational chemical potential $\omega$) known as spin density. In the presence of rotation in the medium, the Euler equation of thermodynamic variables becomes \cite{Florkowski:2017ruc},
 \begin{equation}
     \varepsilon + P = sT + n\mu + \rm{w}\omega.
 \end{equation}

Therefore the new spin density $\rm{w}$ can be calculated as,
 \begin{equation}
     \label{rotdensity}
     \begin{split}
         \rm{w} &= \frac{\partial P}{\partial \omega}\bigg{|}_{T,\mu} = \frac{g_i}{2\pi^2} \int_{0}^{\infty} p^2 dp\ \frac{p^2}{3E_i} \frac{e^{ ({\bf p} \cdot \bf v)/T }}{\exp[(E_i-\mu_i)/T]\pm 1}\\
         & \times \Big[\frac{1}{T}px\ \chiomt + \frac{\partial}{\partial \omega}\chiomt \Big].
     \end{split}
 \end{equation}
Moreover, the $n^{th}$-order susceptibilities of conserved charges can be calculated from the relation,
\begin{equation}
    \label{susept}
    \chi_{x}^{n} = \frac{\partial^{n}(P/T^{4})}{\partial(\frac{\mu_{x}}{T})^{n}},
\end{equation}
where the corresponding conserved charges, such as the baryon number, electric charge, and strangeness number, are represented by the letter x.

The Ideal HRG model does not include interactions among the hadrons and, therefore, is unable to explain different thermodynamic quantities estimated from lQCD at high temperatures and baryon densities. Now to introduce the interaction in the medium, we start with the van der Waals equation, which, in canonical ensemble representation, can be written as \cite{Samanta:2017yhh, W.Greiner123Stocker}
 \begin{equation}
\label{vdweq}
    \Bigg( P + \bigg(\frac{N}{V}\bigg)^{2}a\Bigg)\big(V-Nb\big) =  N T
\end{equation}
where the VDW parameters $a$ and $b$, both positive, describe the attractive and repulsive interactions, respectively, among the hadrons. P, V, T, and N, respectively, stand for pressure, volume, temperature, and the number of particles in the system.

Writing number density, $n \equiv N/V,$ the above equation can be simplified as
\begin{equation}
\label{vdweq2}
    P(T,n) = \frac{nT}{1-bn}- an^{2},
\end{equation}

The two terms in the above equation represent the correction factor to the ideal case due to repulsion and attraction separately. The excluded volume correction, or the correction for repulsive interactions, is incorporated in the first term by changing the total volume V to an effective volume that is accessible to particles using the appropriate volume parameter $b = 16\pi r^3/3$, where $r$ is the particle's hardcore radius. In contrast, the second term accounts for the attractive interactions between particles. For $a = 0$, Eq. (\ref{vdweq2}) reduces to the EVHRG equation of state, where only repulsive interactions are included. And for both $a = 0$, $\&$ $b = 0$, it reduces to the ideal HRG.

This method is then applied to the GCE, where the VDW equation of state takes the form \cite{Samanta:2017yhh,Vovchenko:2015vxa,Vovchenko:2015pya} 
\begin{equation}
\label{vdwp}
    P(T,\mu) = P^{id}(T,\mu^{*}) - an^{2}(T,\mu),
\end{equation}
where, $P(T,\mu)$ is the VDW pressure and reduces to ideal one, $P^{id}(T,\mu)$ when there is no interaction. The particle number density of the VDW hadron gas, n(T,$\mu$) is given by
\begin{equation}
\label{vdwn}
    n(T,\mu) = \frac{\sum_{i}n_{i}^{id}(T,\mu^{*})}{1+b\sum_{i}n_{i}^{id}(T,\mu^{*})}.
\end{equation}
Here, $i$ runs over all hadrons and resonances in the interacting medium, and $\mu^{*}$ is the modified chemical potential given by 
\begin{equation}
\label{mustar1}
    \mu^{*} = \mu - bP(T,\mu) - abn^{2}(T,\mu) + 2an(T,\mu).
\end{equation}
Using Eq. (\ref{vdwp}), the $\mu^{*}$ can also be written as
\begin{equation}
\label{mustar2}
        \mu^{*} = \mu - \frac{bn(T,\mu)T}{1-bn(T,\mu)} + 2an(T,\mu).
\end{equation}
Additional thermodynamical variables like energy density $\varepsilon(T,\mu)$ and entropy density $s(T,\mu)$ can now be calculated as, 
\begin{equation}
\label{vdwe}
\varepsilon(T,\mu) = \frac{\sum_{i}\epsilon_{i}^{id}(T,\mu^{*})}{1+b\sum_{i}n_{i}^{id}(T,\mu^{*})} - an^{2}(T,\mu).
\end{equation}
\begin{equation}
\label{vdws}
s(T,\mu) = \frac{s^{id}(T,\mu^{*})}{1+bn^{id}(T,\mu^{*})},
\end{equation}

As formulated initially, the VDWHRG model includes interactions confined to all pairings of baryons or anti-baryons \cite{Samanta:2017yhh, Vovchenko:2015vxa, Vovchenko:2015pya, Vovchenko:2016rkn}. Considering the fact that annihilation processes dominate short-range interactions between baryon-antibaryon pairs, the interaction between them was neglected \cite{Andronic2012, Vovchenko:2016rkn}. Previously, all meson-related interactions, such as meson-meson or meson-(anti)baryon interactions, are neglected as the inclusion of these interactions suppresses the thermodynamic quantities in the crossover region at vanishing baryochemical potential in comparison with LQCD data \cite{Vovchenko:2016rkn}. However, by assuming a hard-core radius $r_M$ for mesons, a more realistic formalism that takes into account meson-meson interactions was developed by selecting the VDW parameters that best fit the LQCD data~\cite{Sarkar:2018mbk}.
As a result, the VDWHRG model's total pressure is expressed as \cite{Samanta:2017yhh, Pradhan:2022gbm, Vovchenko:2015vxa, Vovchenko:2015pya, Vovchenko:2016rkn, Sarkar:2018mbk} 

\begin{equation}
\label{finalp}
P(T,\mu) = P_{M}(T,\mu) + P_{B}(T,\mu) + P_{\bar{B}}(T,\mu),
\end{equation}
where the pressure contributions made by mesons and (anti)baryons, respectively, are denoted by $P_{M}(T,\mu),  P_{B(\bar B)}(T,\mu)$, and are given by,
\begin{equation}
\label{mesonp}
P_{M}(T,\mu) = \sum_{i\in M}P_{i}^{id}(T,\mu^{*M}),       
\end{equation}
\begin{equation}
\label{baryonp}
P_{B}(T,\mu) = \sum_{i\in B}P_{i}^{id}(T,\mu^{*B})-an^{2}_{B}(T,\mu),
\end{equation}
\begin{equation}
\label{antip}
P_{\bar{B}}(T,\mu) = \sum_{i\in \bar{B}}P_{i}^{id}(T,\mu^{*\bar{B}})-an^{2}_{\bar{B}}(T,\mu).
\end{equation}
Here, mesons, baryons, and anti-baryons are each represented by $M$, $B$, and $\bar B$. Due to the excluded volume correction, mesons have a modified chemical potential of $\mu^{*M}$, whereas baryons and anti-baryons have modified chemical potentials of $\mu^{*B}$ and $\mu^{*\bar B}$, respectively, as a result of VDW interactions \cite{Sarkar:2018mbk}. Taking a simple case of vanishing electric charge and strangeness chemical potentials, where $\mu_{Q} = \mu_{S} = 0$, the modified chemical potential for mesons and (anti)baryons can be determined from Eq.~\ref{mub} and Eq.~\ref{mustar1} as; \begin{equation}
\label{mumeson}
\mu^{*M} = -bP_{M}(T,\mu),
\end{equation}
\begin{equation}
\label{mubaryon}
\mu^{*B(\bar B)} = \mu_{B(\bar B)}-bP_{B(\bar B)}(T,\mu)-abn^{2}_{B(\bar B)}+2an_{B(\bar B)},
\end{equation}
where $n_{M}$, $n_{B}$ and $n_{\bar B}$ are the modified number densities of mesons, baryons, and anti-baryons, respectively, which are given by
\begin{equation}
\label{nmeson}
    n_{M}(T,\mu) = \frac{\sum_{i\in M}n_{i}^{id}(T,\mu^{*M})}{1+b\sum_{i\in M}n_{i}^{id}(T,\mu^{*M})},
\end{equation}
\begin{equation}
\label{nbaryon}
    n_{B(\bar B)}(T,\mu) = \frac{\sum_{i\in B(\bar B)}n_{i}^{id}(T,\mu^{*B(\bar B)})}{1+b\sum_{i\in B(\bar B)}n_{i}^{id}(T,\mu^{*B(\bar B)})}.
\end{equation}

There are different approaches to estimate the VDW parameters. They can be obtained by reproducing the ground state of the nuclear matter \cite{Vovchenko:2015vxa}. Alternatively, one can obtain the parameters by fitting lattice QCD results for different thermodynamic quantities \cite{Samanta:2017yhh, Sarkar:2018mbk}. The parameters in the model are now set to $a=0.926$ GeV fm$^{3}$ and $b=(16/3)\pi r^3$, where the $r$ being the hardcore radius of each hadron, given as $r_{M}=0.2$ fm for mesons, and $r_{B,(\bar{B})}=0.62$ fm, for (anti)baryons~\cite{Sarkar:2018mbk}. Using this information, we now proceed to estimate various thermodynamic quantities in a rotating hadron resonance gas with VDW interactions.

\begin{figure*}[ht!]
\begin{center}
\includegraphics[scale = 0.29]{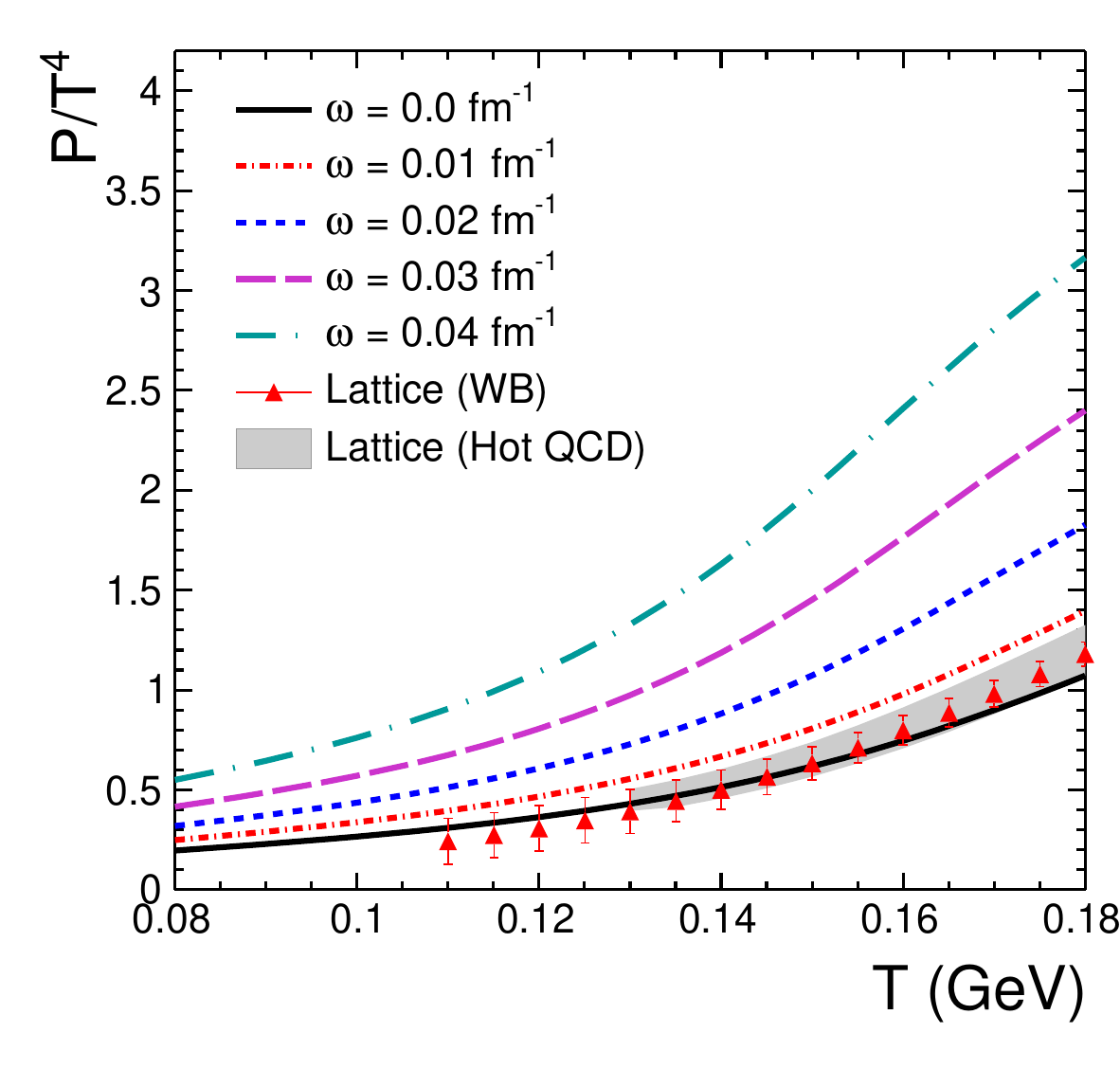}
\includegraphics[scale = 0.29]{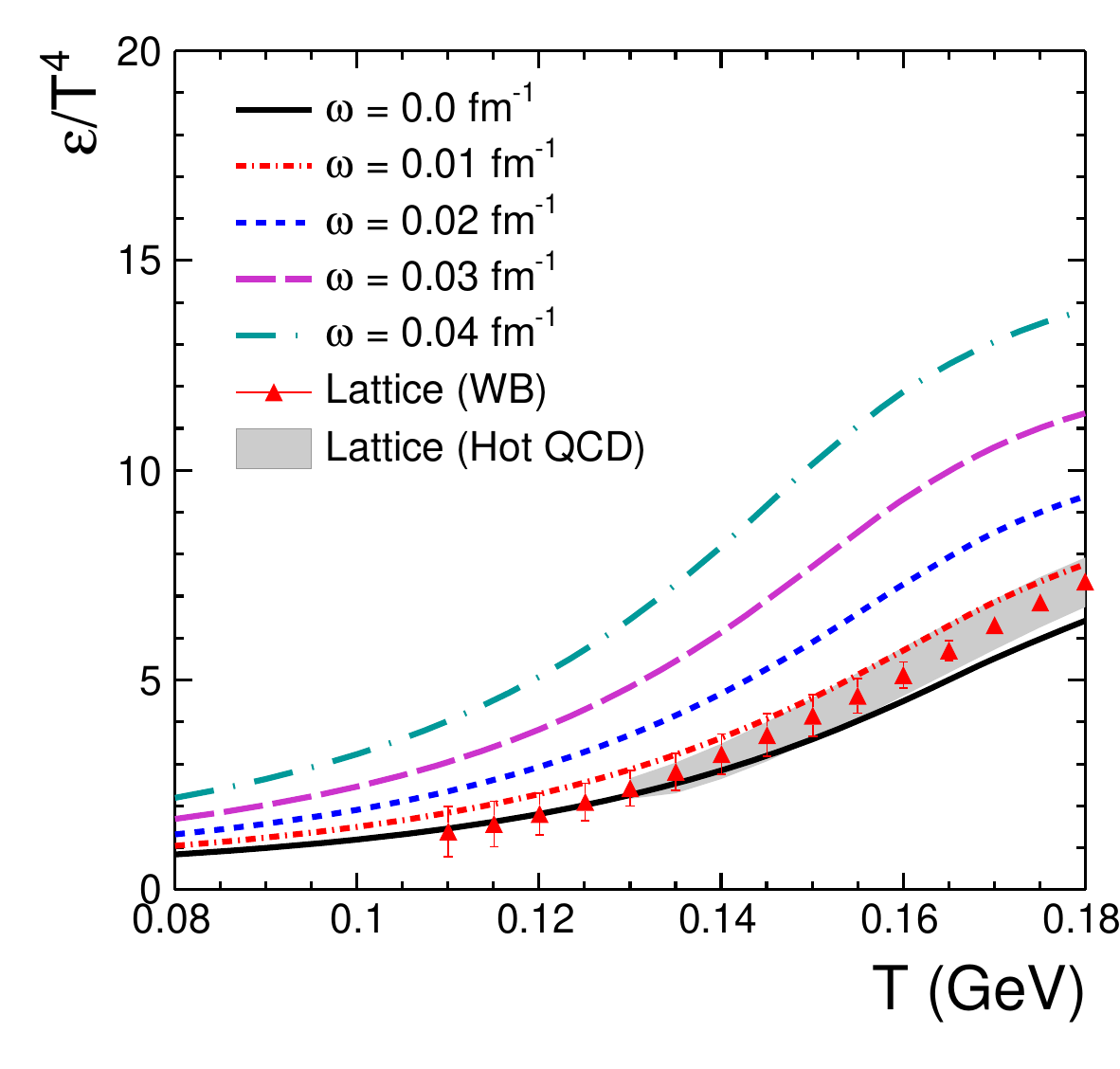}
\includegraphics[scale = 0.29]{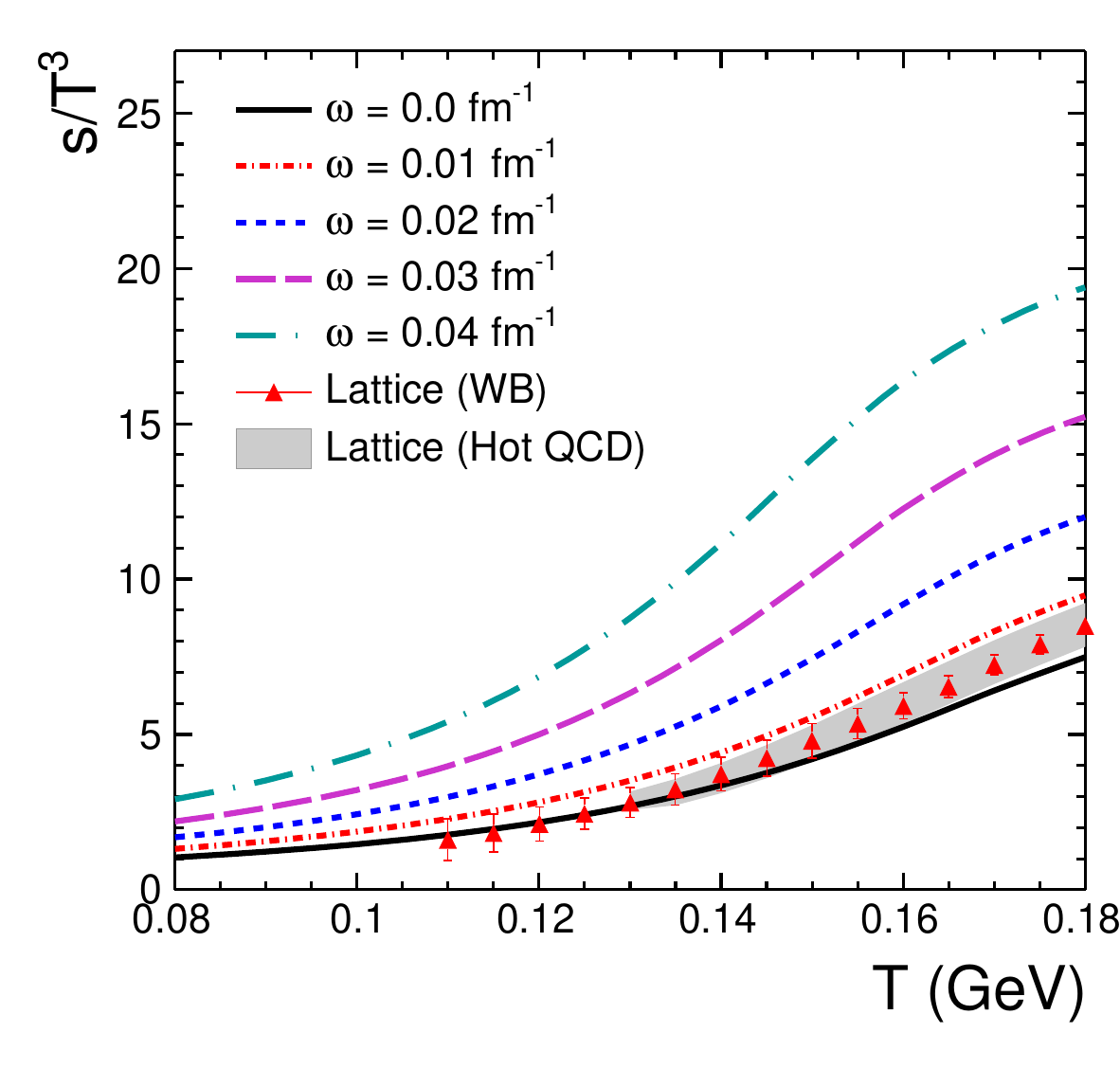}
\includegraphics[scale = 0.29]{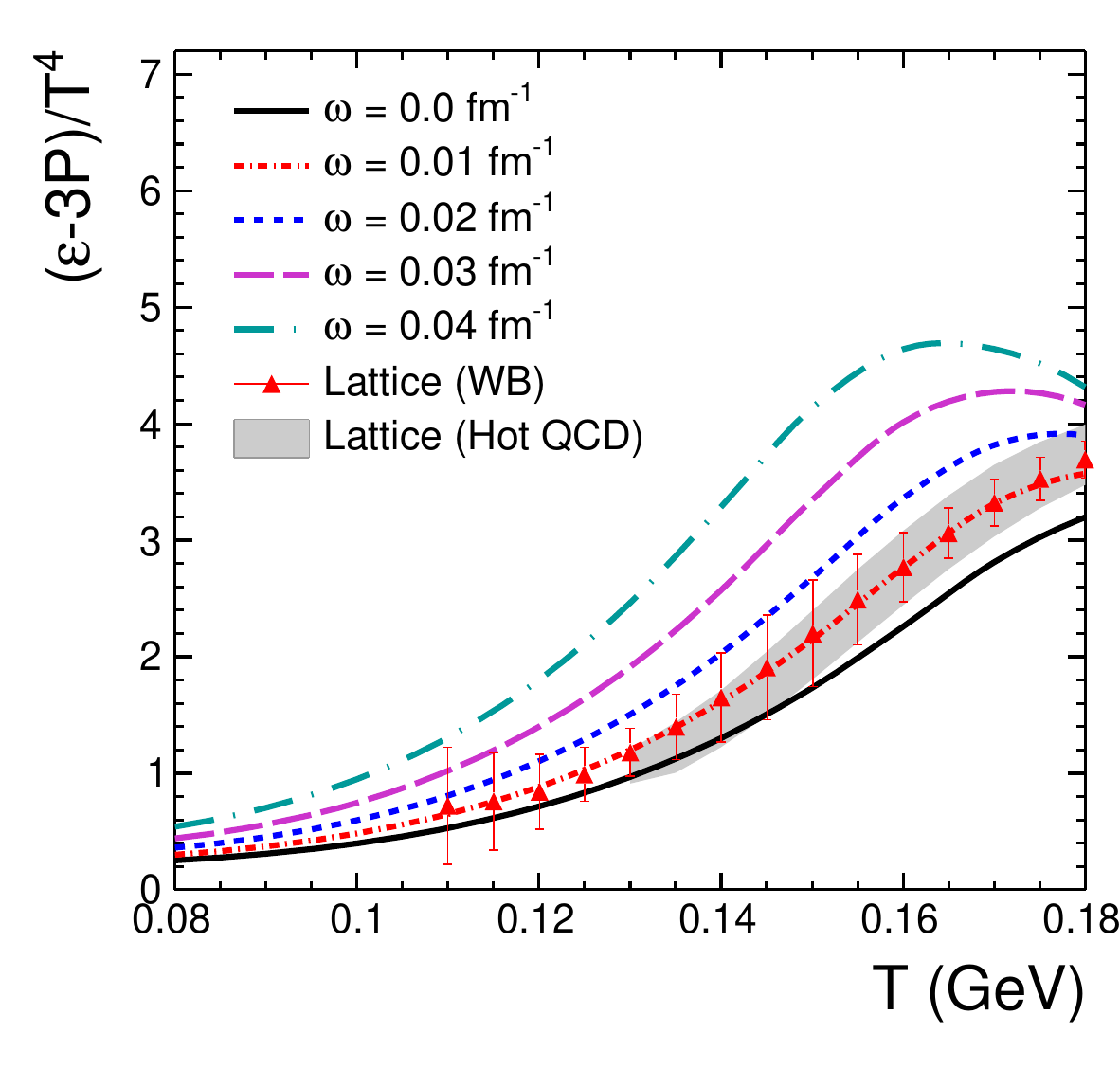}
\includegraphics[scale = 0.29]{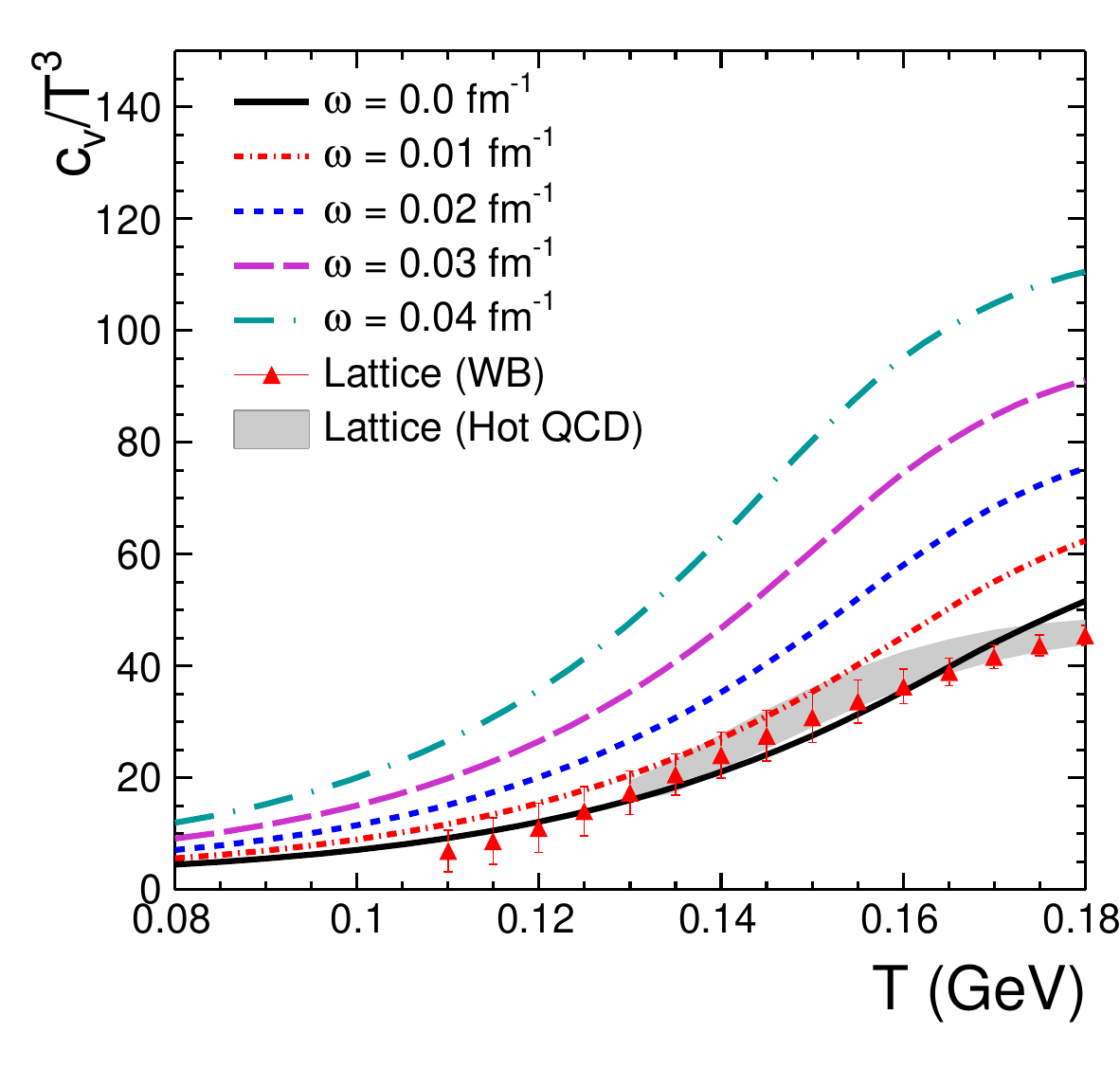}
\includegraphics[scale = 0.29]{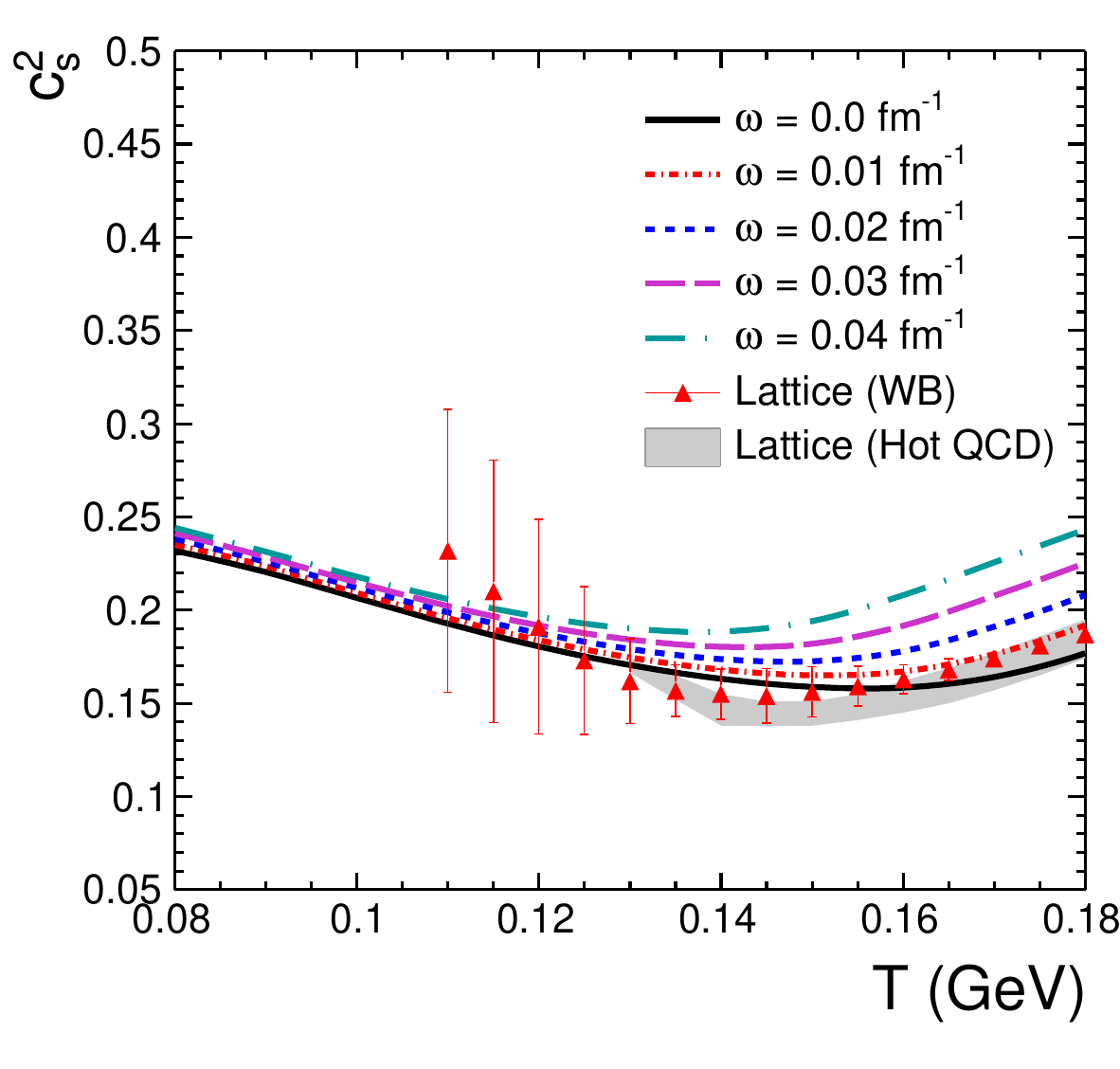}
\caption{(Color Online) Variation of different thermodynamical quantities as functions of temperature for $\mu_{\rm B}$ = 0 GeV, and for certain values of rotation chemical potentials. The red triangles are the Wuppertal-Budapest lattice QCD data \cite{WBcoll}, and the shaded region are the HotQCD lattice data \cite{HotQCD}.}
\label{fig1}
\end{center}
\end{figure*}

\section{Results and Discussion}
\label{results}

We explore the effect of rotation in the hadron gas by considering an interacting hadron gas model, namely, the VDWHRG model, with attractive and repulsive interactions between the hadrons. The model takes into the contributions of all hadrons and resonances up to a mass cut-off of 2.25 GeV available in the particle data group~\cite{PDG2016}. The van der Waals parameters are obtained by fitting thermodynamic quantities like energy density and pressure in the VDWHRG model to the available lattice QCD data \cite{Sarkar:2018mbk}. Notably, the van der Waals parameters should, in principle, change with respect to a change in rotation. However, as it is non-trivial to have $a$ and $b$ as functions of $\omega$, we neglect the dependency in the current study. We then estimate various thermodynamic quantities at finite rotation by taking the obtained $a$ and $b$ values from the fitting. It is also noteworthy to mention that the thermodynamic quantities like pressure in a rotating medium may not be uniform and can have components parallel and perpendicular to the angular momentum vector. However, for simplicity, the formalism here considers all the thermodynamic observables to be isotropic in nature throughout the medium, as done in several earlier works \cite{Wang:2018sur, Fujimoto:2021xix, Mukherjee:2023qvq, Mukherjee:2023ijv}.

\begin{figure}[ht!]
\begin{center}
\includegraphics[scale = 0.44]{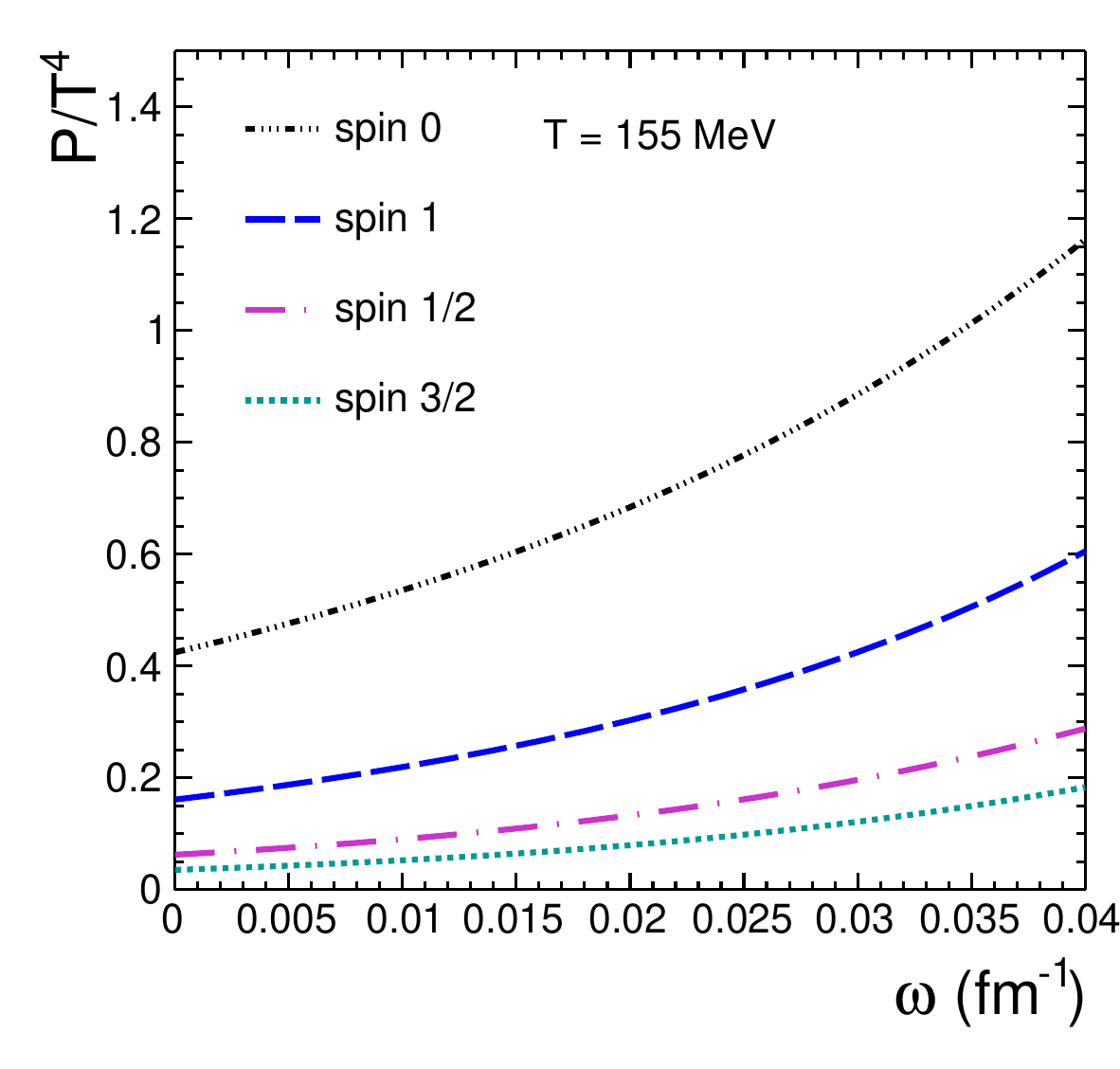}
\caption{(Color Online) Pressure for different spin particles as a function of $\omega$ at a constant temperature, $T$ = 155 MeV.}
\label{fig22}
\end{center}
\end{figure}

Fig.~\ref{fig1} shows the variation of $P/T^{4}$, $\varepsilon/T^{4}$, $s/T^{3}$, $(\varepsilon-3P)/T^{4}$, $c_{v}/T^{3}$, and $c_{\rm s}^{2}$ with temperature at zero baryochemical potential for certain values of $\omega$. The red triangles are the lQCD results from the Wuppertal-Budapest collaboration \cite{WBcoll}, and the shaded region shows the lattice results from Hot QCD collaboration \cite{HotQCD}, at $\mu_{B} = 0.0$ GeV. All the calculations are done for $\mu_{B}$ = 0.0 GeV. Our results at $\omega = 0~ \rm fm^{-1}$, represented by a solid black line, are in good agreement with the lQCD estimations. From the upper-left panel of fig.\ref{fig1}, we observe that $P/T^{4}$ increases with temperature for all $\omega$ values. At a given temperature, $P/T^{4}$ is higher for a higher value of $\omega$. Similar trends can be observed in $\varepsilon/T^{4}$, $s/T^{3}$, $(\varepsilon-3P)/T^{4}$ as well as in $c_{v}/T^{3}$ plots. However, the slopes of the spectra differ for each observable. In the trace anomaly plot, we observe a peak that shifts towards low temperatures with an increase in $\omega$. This peak signifies the conformal symmetry breaking at which the constituent particles become massless. The behavior of $c_{s}^{2}$ is also crucial to understand the phase transition region. In VDWHRG, there appears a minimum in $c_{s}^{2}$, which is in agreement with lQCD, and this minimum can be interpreted as a signature of the transition from hadrons to quark degrees of freedom. By increasing the value of rotational chemical potential, the minima shift towards the lower temperature regime, suggesting that the phase transition temperature decreases in the presence of rotation. 

In order to understand the effect of spin and rotation on the basic thermodynamic quantities, we plot the scaled pressure as a function of $\omega$ for various spin particles in fig.~\ref{fig22}. The temperature is taken to be constant at $T$ = 155 MeV. We observe that the contribution to pressure is dominated by the spin-0 particles, followed by the spin-1, spin-1/2, and spin-3/2 particles. This is due to the fact that the contribution to pressure comes mainly from the Boltzmann factor in the distribution function. Thus, the lesser massive particles will dominantly contribute. The spin-0 particles, which consist of pions, kaons, etc., contribute the most, followed by vector mesons such as $\rho$ and $\phi$. Finally, the spin-1/2 baryons, such as protons and neutrons, and spin-3/2 baryons, such as $\Sigma^*$, $\Xi^*$, will contribute to the pressure, respectively. With the increase in angular velocity, the contribution to pressure increases from all spin particles for the same reason as already mentioned.

\begin{figure}[ht!]
\begin{center}
\includegraphics[scale = 0.44]{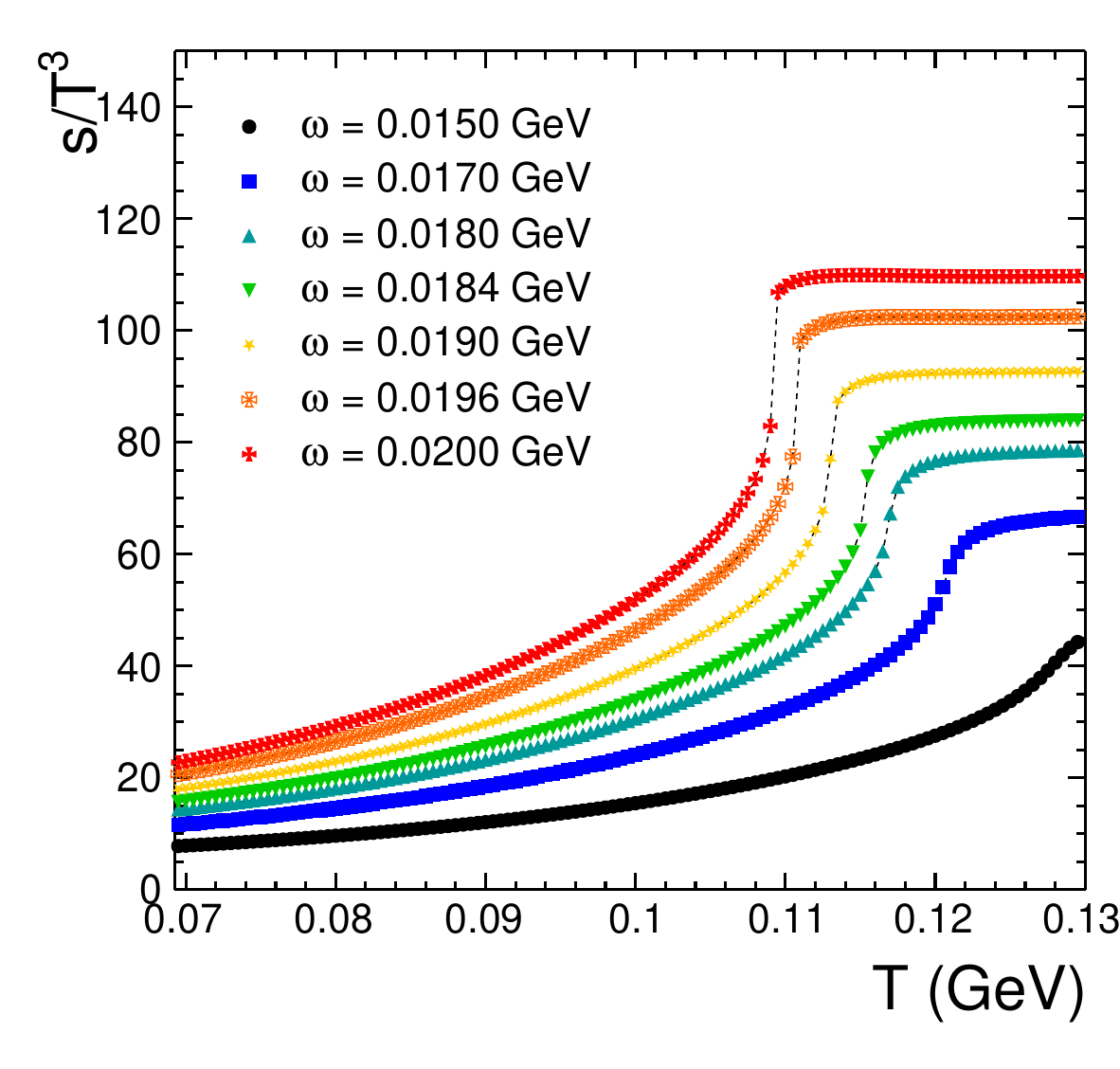}

\caption{(Color Online) Variation of scaled entropy density for $\mu_B = 0$ GeV at low temperature and higher angular velocity values is shown.  }
\label{fig2}
\end{center}
\end{figure}


\begin{figure}[ht!]
\begin{center}
\includegraphics[scale = 0.44]{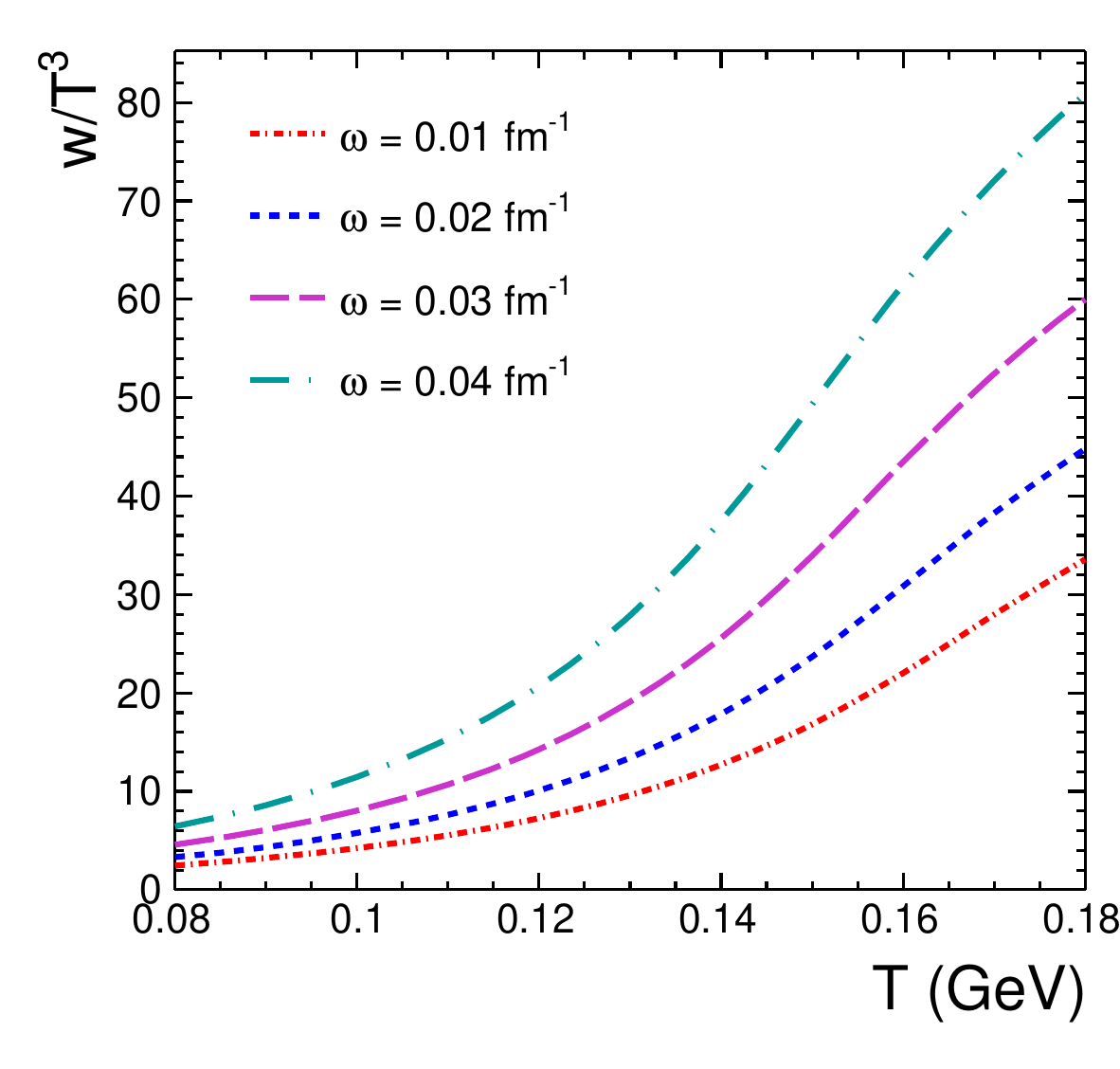}
\caption{(Color Online) Scaled net spin density (w = $\frac{\partial P}{\partial \omega}$) as a function of temperature for different values of $\omega$.}
\label{fig3}
\end{center}
\end{figure}
\begin{figure*}[ht!]
\begin{center}
\includegraphics[scale = 0.29]{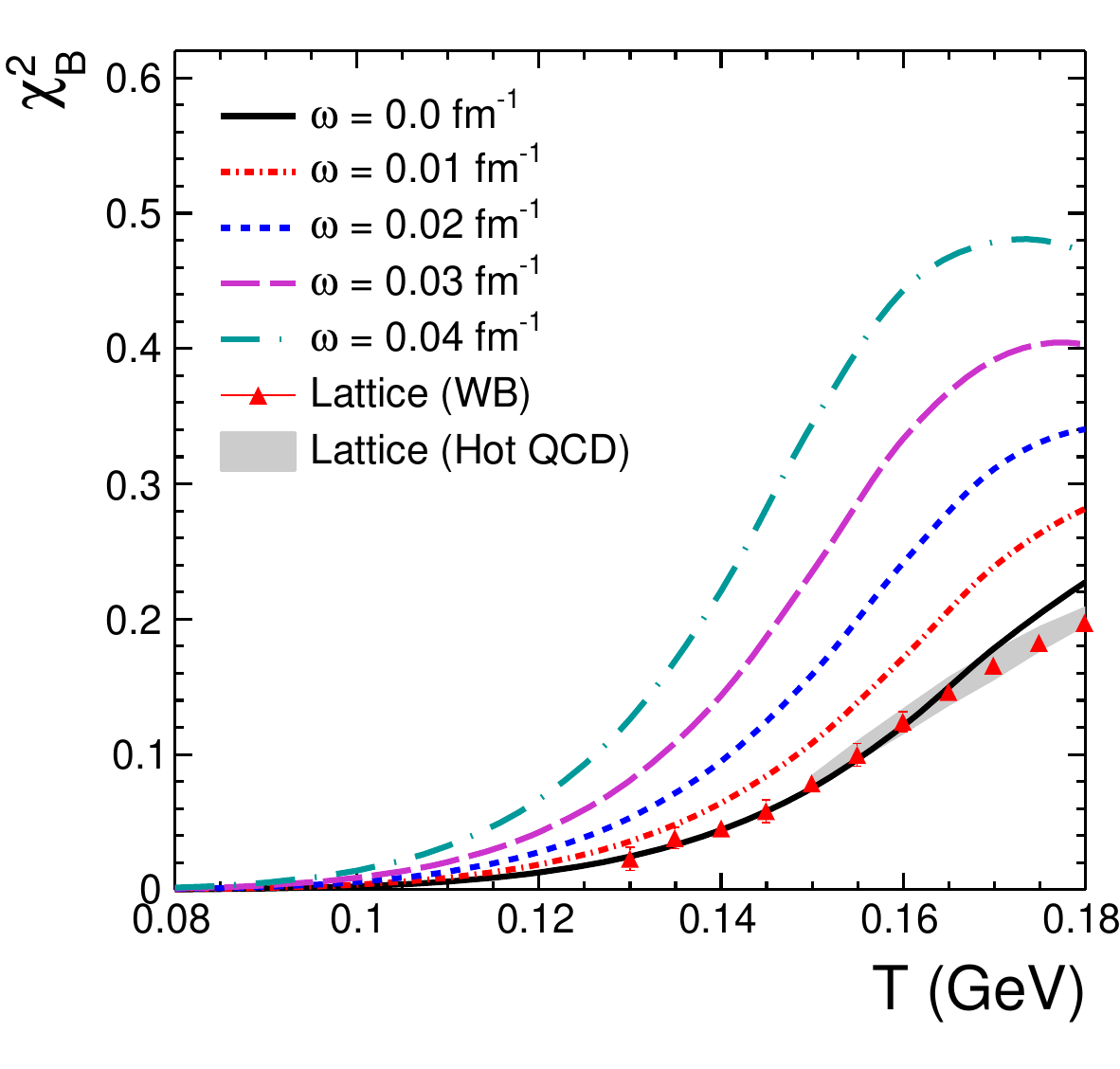}
\includegraphics[scale = 0.29]{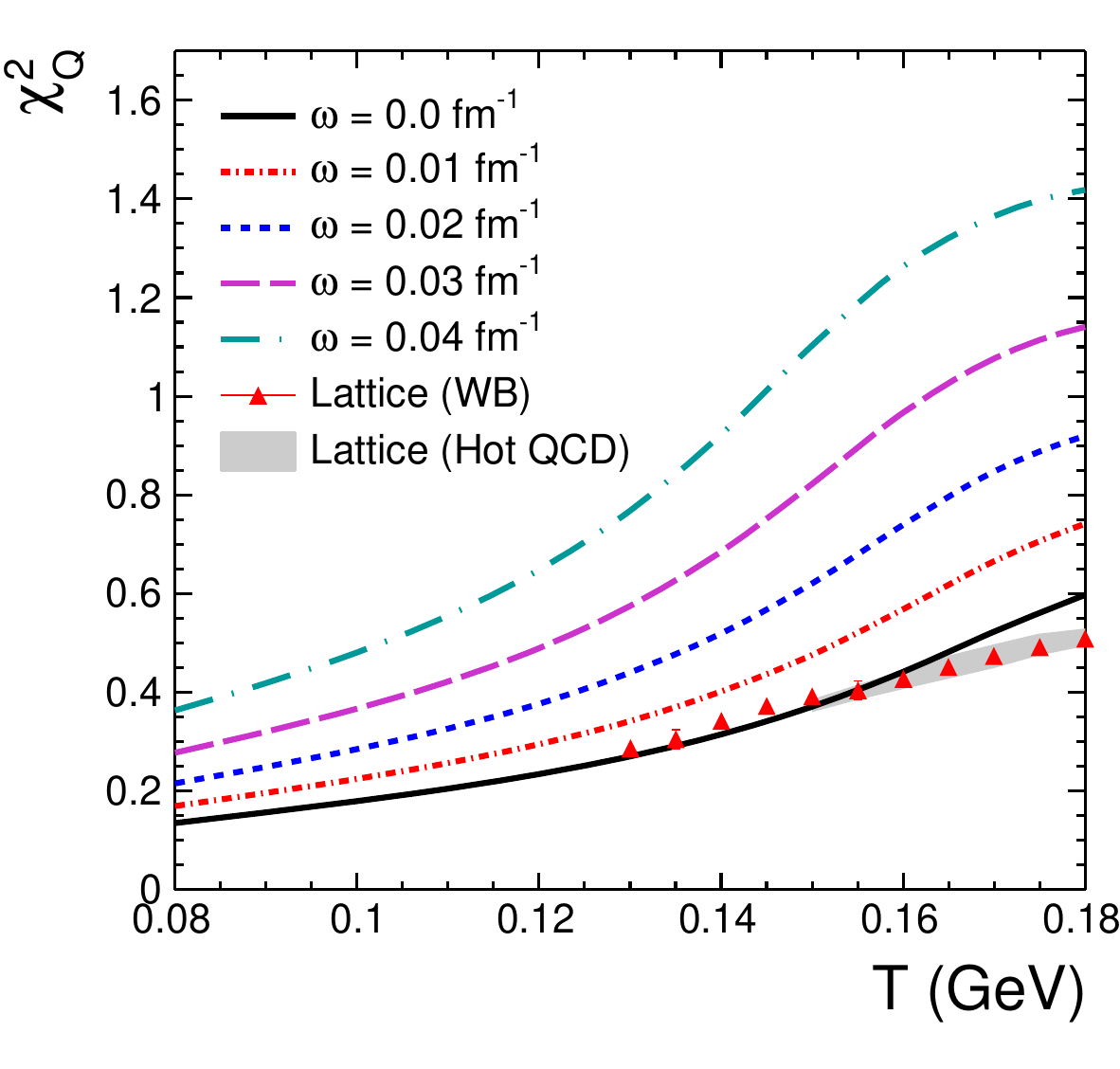}
\includegraphics[scale = 0.29]{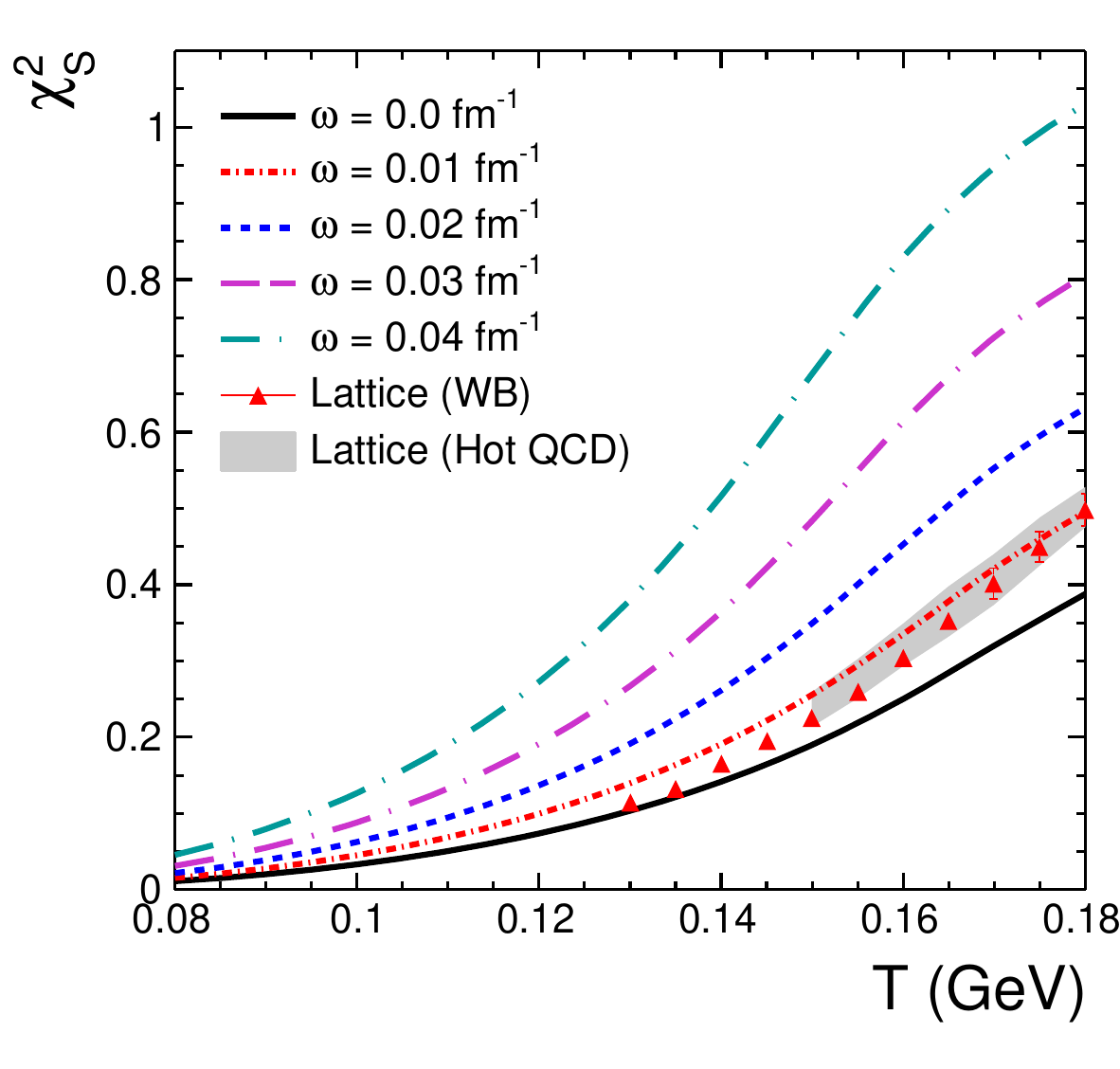}
\caption{(Color Online) The baryon number susceptibility (left panel), charge susceptibility (middle panel), and strangeness susceptibility (right panel) as functions of temperature for different values of $\omega$.}
\label{fig4}
\end{center}
\end{figure*}

 In addition to the speed of sound, the entropy density, number density, etc., are observables that show discontinuities at the first-order phase transition. 
 Our study deals with the van der Waals interaction in the hadronic phase. Therefore a liquid-gas phase transition is expected at $T-\mu_{B}$ plane, which is estimated in various works \cite{Vovchenko:2015pya, Samanta:2017yhh, Sarkar:2018mbk} with different van der Waals parameters. Since rotation adds another chemical potential to the system, it is useful to see if the angular velocity alone can lead to a phase transition. Fig. \ref{fig2} shows the behaviour of scaled entropy density in the $T-\omega$ phase space with $\mu_B = 0.0$ GeV. Here, the angular velocity ($\omega$) is taken in the units of GeV to take into account small iterations. The temperature is taken at an interval of 0.5 MeV for calculation. One can observe a smooth trend of the scaled entropy density at high temperatures and low rotational chemical potential. However, the smooth curve at comparatively low $\omega$ starts changing its shape as one approaches high $\omega$ values. At around $T \simeq 113$ MeV, a discontinuity appears for $\omega \simeq 0.019$ GeV. 
 A clear first-order phase transition is observed as one approaches higher $\omega$ values. This suggests that the rotation has the same effect in achieving a liquid-gas phase transition as the baryon chemical potential~\cite{Samanta:2017yhh}. Therefore, a hadron gas can be liquified either by increasing baryon density and lowering the temperature or by increasing the angular velocity while decreasing the temperature of the gas. Compared to Ref.~\cite{Sarkar:2018mbk}, where the critical point is around $T = 65$ MeV for $T-\mu_{B}$ plane, here the temperature for the critical point is higher, though the VDW parameters are the same. This shows that the phase transition in the presence of a rotational chemical potential appears more quickly than that for the baryochemical potential case. In the holographic QCD approach, a similar effect was found, where the deconfinement critical temperature decreases with an increase in rotation \cite{Chen:2020ath}. However, the recent lattice QCD results \cite{Braguta:2021jgn} show that the deconfinement temperature increases with rotation. The discrepancy between the lattice results and other effective models may be due to the absence of non-perturbative gluonic effects in the effective model calculations. It is to be noted that the authors in Ref. \cite{Fujimoto:2021xix} have used a parametrized relation equating the thermodynamic pressure of hadrons to that of quark-gluon gas to estimate the deconfinement transition temperature, $T_c$, and it is observed that the $T_c$ decreases with the rotation. In our work, we observe similar effects as the effective models, but in the case of liquid-gas phase transition, by studying the $s/T^3$ variable for different values of $\mu_B$ and $\omega$, where the $T_c$ decreases with an increase in rotation.

Fig. \ref{fig3} shows the temperature dependence of normalized dimensionless spin density estimated in the VDWHRG model using Eq.~\ref{rotdensity}. Similar to number density, which can be defined as the change in pressure (or free energy) with respect to chemical potential, spin density can also be defined as the change in the pressure as a function of rotational chemical potential. The net spin density in a system is defined as the density of hadrons of positive spin minus the density of hadrons of negative spin. It is observed that much like other thermodynamic densities, such as number density and entropy density, spin density also increases with an increase in temperature. Moreover, at a particular temperature, the value of spin density increases with increased rotational chemical potential. 
\begin{figure}[ht!]
\begin{center}
\includegraphics[scale = 0.44]{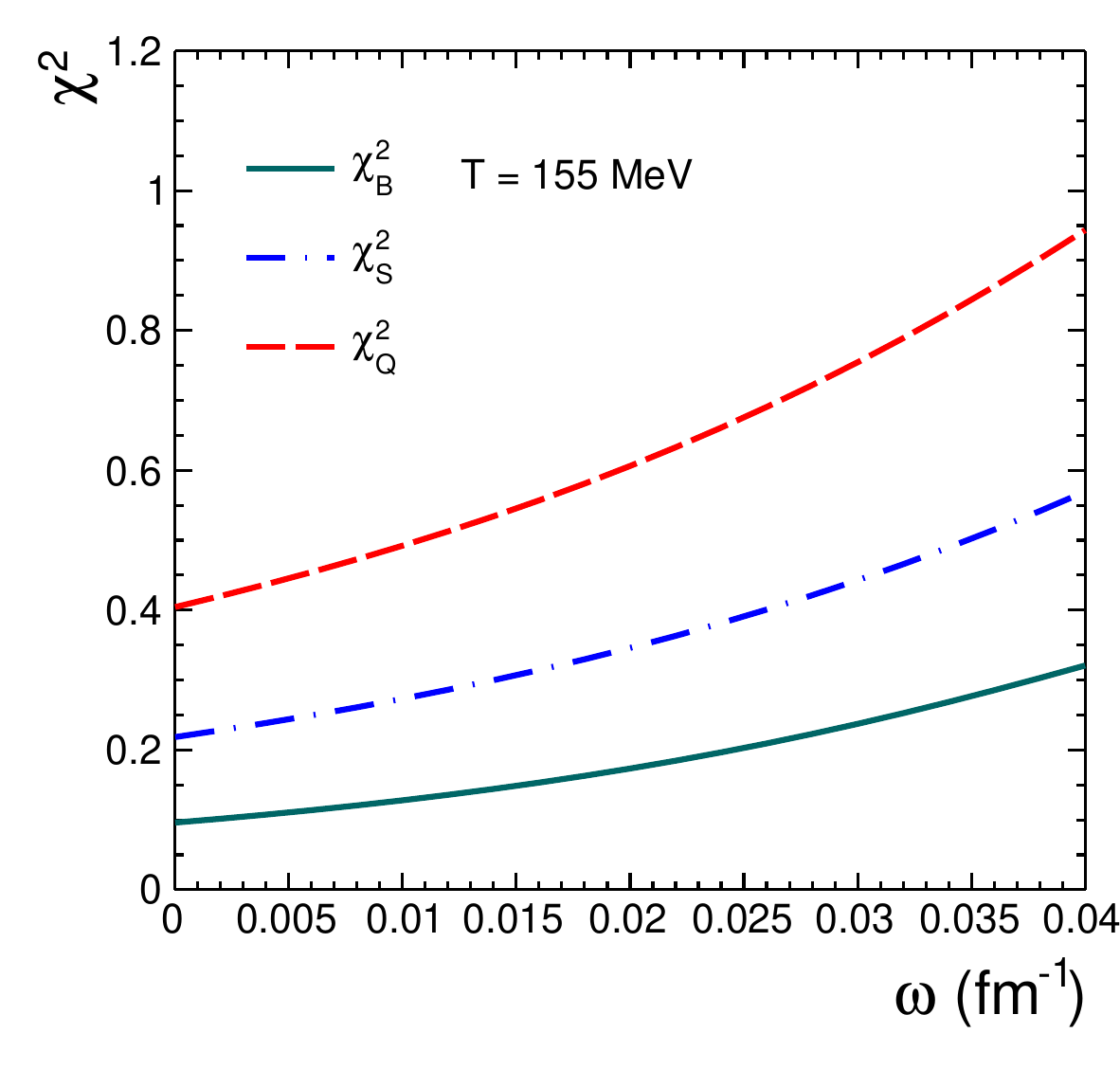}
\caption{(Color Online) Various susceptibilities as functions of $\omega$ at a constant temperature, $T$ = 155 MeV.}
\label{fig5}
\end{center}
\end{figure}

We also estimate the susceptibilities of various conserved quantities to show their dependence on rotational chemical potential. Fluctuations of conserved charges like net baryon density, electric charge, and strangeness are essential probes for hadronization and can help us locate the phase boundary. Large fluctuations in these quantities are one of the essential signatures of the critical endpoint. Since the rotation can affect the phase transition and hence the critical point, it is essential to see its effect on the fluctuations of different conserved charges. We have used Eq. \ref{susept} to calculate the second-order susceptibilities of different conserved quantities. Fig. \ref{fig4} shows the temperature dependence of second-order fluctuations of conserved quantities, namely, baryon density, electric charge, and strangeness, respectively, from left to right. The red triangles are the results of lQCD calculations from the Wuppertal-Budapest collaboration \cite{WBcoll} whereas the shaded region represents the lQCD results of HotQCD collaboration \cite{HotQCD}. The solid black line, calculated in the VDWHRG model at zero baryochemical potential and zero rotation, agrees with lattice results for baryon and charge susceptibility. However, the results for strangeness susceptibility are slightly suppressed than those of lattice results. The fluctuations in every conserved quantity increase with an increase in rotational chemical potential. It is observed that for higher $\omega$, a more prominent peak appears in the case of baryon density fluctuations. However, the trends seem to be saturated in the case of charge susceptibilities, and a monotonic increasing behavior is observed for strangeness fluctuations within the range of studied $\omega$. Fig. \ref{fig5} shows the variation of all three susceptibilities as a function of rotational chemical potential at a fixed temperature $T = 155$ MeV. It is observed that all the susceptibilities increase with $\omega$ almost in a similar manner. 

\section{Summary}
\label{sum}
In this work, we estimate the effect of rotation on the thermodynamic properties of an interacting hadron resonance gas. We observe that rotation has a similar effect on the thermodynamic properties as the baryon chemical potential. The rotational chemical potential enhances all observables like pressure, energy density, entropy density, etc. We also observe that the rotation in a system could lead to a first-order liquid-gas phase transition, although the initial angular momentum required for it would be so high that within LHC energy, it may not be possible. In addition,  we estimate the spin density associated with the rotational chemical potential and its behaviour as a function of temperature. The effect of rotation on fluctuations in conserved quantities is also explored, and one can find that it enhances the second-order fluctuations in all conserved quantities. In view of our study, we must pay attention to the effect of rotation produced in a non-central heavy-ion collision while studying the particle dynamics and the thermodynamics of the system.

Recent studies focusing on vorticity and polarization in the medium formed in ultra-relativistic collisions lead us to an exciting pathway. As the scientific community shifts its attention to rotational dynamics in the evolving QCD medium, myriad unique consequences can be unravelled. Moreover, it will be interesting to see the results of the lattice calculation by taking care of the rotation into the system.
\section*{Acknowledgement}

K.K.P. and B.S. acknowledge the financial aid from UGC and CSIR, Government of India, respectively. The authors gratefully acknowledge the DAE-DST, Government of India funding under the mega-science project “Indian Participation in
the ALICE experiment at CERN” bearing Project No.
SR/MF/PS-02/2021-IITI (E-37123).

\end{document}